\documentclass[aps,nofootinbib,showpacs,showkeys,final,balancelastpage,superscriptaddress]{revtex4}
\usepackage{amssymb}
\usepackage{amsmath}
\usepackage{amsbsy}
\usepackage{amsfonts}

\DeclareFontFamily{OT1}{pzc}{}
\DeclareFontShape{OT1}{pzc}{m}{it}{<-> s * [1.10] pzcmi7t}{}
\DeclareMathAlphabet{\mathpzc}{OT1}{pzc}{m}{it}
\providecommand{\U}[1]{\protect\rule{.1in}{.1in}}
\DeclareFontFamily{OT1}{pzc}{}
\DeclareFontShape{OT1}{pzc}{m}{it}{<-> s * [1.10] pzcmi7t}{}
\DeclareMathAlphabet{\mathpzc}{OT1}{pzc}{m}{it}
\providecommand{\U}[1]{\protect\rule{.1in}{.1in}}
\providecommand{\U}[1]{\protect\rule{.1in}{.1in}}
\providecommand{\U}[1]{\protect\rule{.1in}{.1in}}
\providecommand{\U}[1]{\protect\rule{.1in}{.1in}}
\providecommand{\U}[1]{\protect\rule{.1in}{.1in}}
\input{tcilatex}
\begin{document}

\title{ Nonlinear magnetic response of the magnetized vacuum to applied
electric field }
\author{Dmitry M. \surname{Gitman}}
\affiliation{Instituto de F\'{\i}sica, Universidade de S\~{a}o Paulo, S.P., Brazil}
\email{gitman@dfn.if.usp.br}
\author{Anatoly \surname{E. Shabad}}
\affiliation{P. N. Lebedev Physics Institute, Moscow, Russia.}
\email{shabad@lpi.ru}

\begin{abstract}
We find first nonlinear correction to the field, produced by a static charge
at rest in a background constant magnetic field. It is quadratic in the
charge and purely magnetic. The third-rank polarization tensor - the
nonlinear response function - is written within the local approximation of
the effective action in an otherwise model- and approximation-independent
way within any P-invariant nonlinear electrodynamics, QED included.
\end{abstract}

\maketitle

%\date{\today }

\section{ \ \ \ \ \ \protect\bigskip Introduction}

\bigskip

In Maxwell electrodynamics\textbf{\ }the superposition principle is true,
which reads that electromagnetic fields do not directly interact between
themselves and may be linearly combined independently. This is not the case
in nonlinear electrodynamics, wherein only small electromagnetic fields are
mutually independent.

\bigskip

A popular example of a nonlinear electrodynamics in the vacuum is provided
by the Born-Infeld model \cite{born}, also by a noncommutative $U_{\star}(1)$
gauge theory, in this respect considered, $e.g.,$ in \cite{Stern, AGSV} .
Many issues of nonlinear electrodynamics are thoroughly elaborated in \cite%
{plebanski}. Another, practically most important example is quantum
electrodynamics (QED). The reason why it is nonlinear is that an
electromagnetic field, say a photon, may create virtual electron-positron
pairs that interact with this field itself and/or with any other, "external"
field. This makes a mechanism that lets electromagnetic fields sense each
other.

\bigskip

The well-known nonlinear effect of QED, present already in the vacuum
without any external field, is light-by-light scattering. When taken off the
photon mass shell the corresponding probability amplitude becomes as a
matter of fact responsible for the leading nonlinear (cubic) correction to
the electric Coulomb field \cite{BerLifPit}\footnote{%
The authors are indebted to M.I. Vysotsky, who attracted their attention to
this result.} that can be conveniently written as

\begin{equation}
\mathcal{E}_{\text{nl}}\mathcal{=E}\left( 1-\frac{2\alpha }{45\pi }\left( 
\frac{e\mathcal{E}}{m^{2}}\right) ^{2}\right) .  \label{BerLifPit}
\end{equation}%
Here $\mathcal{E=(}q/4\pi r^{2})$ is the standard Coulomb field\footnote{%
The linear response to the applied charge $q$ due to the vacuum polarization
known as the Uehling-Silber correction to the Coulomb potential \cite%
{BerLifPit} may be also included into $\mathcal{E}$.} in Heaviside units
produced by the point charge $q$ at the distance $r$, while $e$ and $m$ are
the electron charge and mass, $\alpha =(e^{2}/4\pi )=1/137$ is the
fine-structure constant.\ It is generally known, and also seen in this
equation, that in QED the nonlinearity is determined by the ratio of the
electromagnetic field to Schwinger's characteristic value $%
(m^{2}/e)=4.4\cdot 10^{13}$ cgse units that makes 1.3$\cdot 10^{16}$ V/cm
when one measures an electric field, and $4.4\cdot 10^{13}$ G if a magnetic
field is concerned. Electromagnetic fields should be comparable in strength
to these values in order that the interaction between them migt become
essential. The nonlinear correction in (\ref{BerLifPit}) becomes valuable
when one is interested to approach a sufficiently small-sized charge
sufficiently close. Say, to approach the nucleus of a not too heavy atom as
close as a few Fm. On the other hand, electric fields, large in the
Schwinger scale, up to $10^{18}-10^{19}$ V/cm, occur \cite{Co and Usov} at
the surface of strange quark stars \cite{witten}, depending on whether the
matter is in the superconducting state \cite{linares}. For such fields the
vacuum is unstable, and the Schwinger effect of spontaneous
electron-positron pairs by the vacuum becomes already efficient, which
requires a special treatment, see the book \cite{FGS}. We do not consider
the corresponding complications in the present paper, however.

In this paper we are dealing with another nonlinear phenomenon, also
associated with strong electric field, namely the production by it of a
magnetic field: this magneto-electric effect becomes possible if an external
magnetic field is present.

The linear correction to the Coulomb field of a charge due to the vacuum
polarization in a magnetic field was studied earlier \cite{shabus, sad,
mache} with the result that the hydrogen ground energy level saturates \cite%
{shabus, moche} as the magnetic field grows, and that a string is formed 
\cite{shabus}. Some hints were thereby produced for considering \cite%
{simonov} interquark potential in QCD. The nonlinear (purely magnetic)
correction to the field of a charge in a magnetic field to be considered now
for the first time is based on the known fact that in this case not only the
photon-by-photon scattering exists, but also the photon splitting into two
(also two-photon merging into one). The splitting is enhanced by the
strength of the external magnetic field as compared to the vacuum case
above. It was elaborated in theory \cite{adler} and is thought of as being
efficient in a pulsar magnetosphere with the magnetic fields above $10^{12}$
G \cite{harding}, essentially contributing to the electron-positron plasma
production and to the radiation pattern of pulsars. Again, the same as
above, when taken outside the photon mass shell, the corresponding
probability amplitudes become responsible for a nonlinear induction of
time-independent current (and, hence, of the stationary magnetic field) by
static charges or, equivalently, by static electric fields created by them.
The magnetic field produced by a static charge in an external magnetic field
is even (quadratic in the lowest order of nonlinearity) with respect to its
magnitude and linearly diappears with the external field -- in agreement
with the generalized Furry theorem of Ref. \cite{AGSV} that states that the
numbers of electric and magnetic legs in every diagram should be each even.
It also agrees with this theorem in that there are no corrections to the
static electric field in the lowest (second-power) nonlinear order.
Previously, magneto-electric effect was considered in \cite{Stern, AGSV} for
classical noncommutative electrodynamics, and within QED as a linear
response to a static charge by the vacuum filled with external electric and
magnetic fields \cite{shabus2010}.

In the next Section II, for the most general case of a constant and
homogeneous external electromagnetic field, we outline the derivation of
nonlinear Maxwell equations keeping only the first and the second powers of
the electromagnetic field living above that external field background, and
define a notion of a current, nonlinearly induced by a static electric field
(or by a static charge). The nonlinear field equations are served by the
second- and third-rank polarization tensors. In Section III we restrict the
external background to the magnetic-like field, i.e. the one that is purely
magnetic in a class of special Lorentz frames. Then the involved
polarization tensors are given in small-4-momentum limit, called also
infrared or local approximation, in terms of the derivatives of the
effective Lagrange density over the background field invariants, bearing in
mind that in the local approximation this density does not depend upon
space-time derivatives of the background field strength. In Section IV we
are working in a special frame, where the background field is purely
magnetic and the static charge is at rest. We calculate the
nonlinearly-induced current and its magnetic field as expressed through the
static electric field produced by the charge. The limiting cases of very
large and very small background magnetic field are discussed within QED
referring to the one-loop Euler-Heisenberg effective Lagrangian. In
Conclusions the results are resumed, and numerical estimates of the domains
of their applicability are given. Detailed calculations of the second- and
third-rank derivatives of the effective action used in the work are
presented in Appendix within the necessary local approximation.

\section{Nonlinear electromagnetic field equations over a constant field
background}

\bigskip

In QED and in any other U(1)-gauge-invariant nonlinear electrodynamics, the
field equations, when written up to terms, quadratic in the small
electromagnetic field potential $a^{\nu}(x),$ have the form%
\begin{align}
& \left[ \eta_{\rho\nu}\square-\partial^{\rho}\partial^{\nu}\right] a^{\nu
}(x)+\int d^{4}x^{\prime}\Pi_{\rho\nu}(x,x^{\prime})a^{\nu}(x^{\prime })+ 
\notag \\
& +\frac{1}{2}\int d^{4}x^{\prime}d^{4}x^{\prime\prime}\Pi_{\rho\nu\sigma
}(x,x^{\prime}x^{\prime\prime})a^{\nu}(x^{\prime})a^{\sigma}(x^{\prime\prime
})=j_{\rho}(x),  \label{cubic}
\end{align}
where $j_{\rho}(x)$ is a (small) source of the field, Greek indices span the
4-dimensional Minkowski space taking the values 1,2,3,0, the metric tensor
is $\eta_{\rho\nu}=$ diag $(1,1,1,-1)$, and $\square=\nabla^{2}-%
\partial_{0}^{2}$. The second- and the third-rank polarization tensors, $%
\Pi_{\rho\nu}$ and $\Pi_{\rho\nu\sigma},$ here are, in the presence of an
external field potential $A^{\beta}\left( x\right) =\mathcal{A}_{\text{ext}%
}^{\beta}(x),\ $defined as%
\begin{equation}
\Pi_{\mu\tau}(x,x^{\prime})=\left. \frac{\delta^{2}\Gamma}{\delta A^{\mu
}(x)\delta A^{\tau}(x^{\prime})}\right\vert _{A=\mathcal{A}_{\text{ext}}}
\label{Pi}
\end{equation}%
\begin{equation}
\Pi_{\mu\tau\sigma}(x,x^{\prime},x^{\prime\prime})=\left. \frac{\delta
^{3}\Gamma}{\delta A^{\mu}(x)\delta A^{\tau}(x^{\prime})\delta A^{\sigma
}(x^{\prime\prime})}\right\vert _{A=\mathcal{A}_{\text{ext}}}  \label{Pi3}
\end{equation}
in terms of the effective action 
\begin{equation}
\Gamma=\int\mathfrak{L}(z)\mathrm{d}^{4}z,  \label{Gamma}
\end{equation}
the generating functional of all-rank polarization tensors -- the vertex
functions -- known in QED as the Legendre transform of the generating
functional of the Green functions \cite{weinberg}. The parameter of the
power expansion, to which Eq.(\ref{cubic}) provides two lowest terms,
depends on a field scale of a definite dynamical theory. We shall discuss
this issue in Section IV below for QED.

We did not write the zero-power term ($a^{\nu }(x))^{0},$ an external
macroscopic current, in\ equation (\ref{cubic}), because we assumed that the
external field had been subjected to the sourceless field equation 
\begin{equation}
\left. \frac{\delta S}{\delta A^{\beta }\left( y\right) }\right\vert _{A=%
\mathcal{A}_{\text{ext}}}=0,  \label{extfieldeq}
\end{equation}%
where \ 
\begin{equation}
S=\int L(z)\mathrm{d}^{4}z,\text{ \ \ \ }L(z)=-\mathfrak{F}(z)+\mathfrak{L}%
(z)  \label{SL}
\end{equation}%
are the total action and the total Lagrangian, respectively. Here $-%
\mathfrak{F}(z)$ =$(1/4)$ $F_{\mu \nu }F^{\nu \mu }$ is the (free) Maxwell
Lagrangian, $F_{\alpha \beta }(z)=\partial ^{\alpha }A_{\beta }(z)-\partial
^{\beta }A_{\alpha }(z)$ is the field-strength tensor$.$ In\ what follows we
shall only deal with external fields $\mathcal{F}_{\alpha \beta }\mathcal{%
=\partial }^{\alpha }\mathcal{A}_{\beta }^{\text{ext}}\mathcal{-\partial }%
^{\beta }\mathcal{A}_{\alpha }^{\text{ext}},\mathcal{\ }$which are
independent of the 4-coordinate $z_{\mu },$ and with the case where the
effective Lagrangian $\mathfrak{L}(z)$ may depend on $z_{\mu }$\ only
through the field tensor $F_{\alpha \beta }(z)$ and its space-time
derivatives, and not explicitly. The latter property is fulfilled in QED and
will be also assumed for other theories subject to our consideration. Under
this assumption the constant field does satisfy the exact sourceless
nonlinear field equation (\ref{extfieldeq}). To see this, we fulfill the
variational derivative in it 
\begin{equation*}
\left. \frac{\delta S}{\delta A_{\beta }\left( x\right) }\right\vert _{A=%
\mathcal{A}_{\text{ext}}}=2\dsum\limits_{n}\int \left. \frac{\delta S}{%
\delta F_{\alpha \beta }^{\left( n\right) }(z)}\right\vert _{F=\mathcal{F}}%
\frac{\partial }{\partial z^{\alpha }}\delta ^{4(n)}(x-z)\mathrm{d}^{4}z,
\end{equation*}%
where $\left( n\right) $ marks the derivative with respect to any space-time
component. Once the variational derivative $\frac{\delta S}{\delta F_{\alpha
\beta }^{\left( n\right) }(z)},$ when restricted onto the
coordinate-independent fields $F_{\mu \nu }(z)=\mathcal{F}_{\mu \nu },$
cannot depend on $z,$ the integration by parts turns this integral to zero.

The above presentation explains why Eq. (\ref{cubic}) is the field equation
for small electromagnetic perturbations $a^{\beta}(x)=A^{\beta}\left(
x\right) -\mathcal{A}_{\text{ext}}^{\beta}(x)$ over the external field of a
constant field strength, caused by a small external current $j_{\rho}(x)$
and taken to the lowest-power nonlinearity.

Polarization tensors of every rank \ $\Pi_{\mu\tau...\sigma}(x,x^{\prime
},...x^{\prime\prime})$\ \ satisfy the continuity relations with respect to
every argument and every index (the transversality property)\ \ 
\begin{equation}
\frac{\partial}{\partial x_{\tau}^{\prime}}\Pi_{\mu...\tau...\sigma
}(x,...x^{\prime},...x^{\prime\prime})=0,  \label{transvers}
\end{equation}
necessary to provide invariance of every term in\ the expansion of $\ \Gamma$
in powers of the field $a^{\nu}$\ under the gauge transformation of it$.$
Note that this is the primary property of $\ \Gamma$ as a functional given
on field strengths and their space-time derivatives only.

In our case of the external field with space- and time-independent strength
the translational invariance holds true, which fact makes the all-rank
polarization tensors depending on their coordinate differences.

With the definition of the photon propagator $D_{\mu\nu}(x,x^{\prime})$%
\begin{equation}
D_{\mu\nu}^{-1}(x-x^{\prime})=\left[ \eta_{\mu\nu}\square-\partial^{\mu
}\partial^{\nu}\right] \delta^{(4)}(x^{\prime}-x)+\Pi_{\mu\nu}(x-x^{\prime})
\label{propagator}
\end{equation}
the nonlinear field equations (\ref{cubic}) take the form of (the set of)
integral equations

\begin{equation}
a^{\lambda}(x)=\int d^{4}yD^{\lambda\rho}(x-y)j_{\rho}(y)+\int
d^{4}yD^{\lambda\rho}(x-y)j_{\rho}^{\text{nl}}(y),  \label{a}
\end{equation}

\begin{equation}
j_{\mu}^{\text{nl}}(x)=-\frac{1}{2}\int d^{4}yd^{4}u\Pi_{\mu\tau\sigma }^{%
\text{ \ }}(x-u,\text{ }y-u)a^{\tau}(y)a^{\sigma}(u),  \label{nonlincur}
\end{equation}
where we have introduced the notation $j_{\mu}^{\text{nl}}(x)$ for what we
shall be calling "nonlinearly induced current"$.$

Before proceeding, the following explanation seems to be in order. Within
the present approach the electromagnetic field $\ a^{\lambda}(x)$ is not
quantized, this is not needed unless we leave the electromagnetic sector.
The nonlinear equations written in this section are classical and will be
treated classically below in understanding that the effective action is
known. In QED the latter is the final product of quantum theory, obtained by
continual integration over fermions \cite{weinberg}. The effective
Lagrangian and all-rank polarization tensors involved are subject to
approximate quantum calculations and, hence, are functions containing the
Plank constant, electron mass and charge. Available is the effective action
in the local limit referred to in the next section, which is known as the
Euler-Heisenberg action when it is calculated within the approximation of
one electron-positron loop, see \cite{BerLifPit}, and as Ritus action when
it is calculated with two-loop accuracy \cite{ritus}. The second-rank
polarization tensor (\ref{Pi}) was calculated in the one-loop approximation
when the external background is formed by a constant and homogeneous
electromagnetic field of the most general form (when the both its invariants 
$\mathfrak{F}$ and $\mathfrak{G}$ are nonvanishing) in \cite{batalin}.
One-loop diagrams with three photon legs corresponding to the third-rank
tensor (\ref{Pi3}) were calculated both on and off the photon mass shell for
QED with external magnetic-like ($\mathfrak{F>}$ $0,$ $\mathfrak{G=}$ $0$)
and crossed ($\mathfrak{F=G=}$ $0)$ fields in \cite{adler}, and for
charge-asymmetric electron-positron plasma without external field using the
temperature Green function techniques, in \cite{ferinsshab}. The
calculations of Stoneham in \cite{adler} might become a basis for extending
the results of the next sections beyound the local approximation used in
getting them, but they are overcomplicated and not well-structured, so we
leave this extension for future. In the next sections we stick to the
general form of the effective Lagrangian and refer to its specific
Euler-Heisenberg form only at the very last steps for getting numerical
estimates.

\section{Local limit}

From now on we shall restrict ourselves only to slowly varying fields $%
a^{\lambda }(x)$ and, correspondingly, to consideration of the sources $%
j_{\rho }(y)$ that give rise to such fields via equations (\ref{a}), (\ref%
{nonlincur}). To this end we may take the effective action in the local
limit, where the field derivatives are disregarded from this functional.
This is equivalent to going to the infrared asymptotic limit in the second-
and third-rank polarization operators, i.e. to keeping, respectively, only
the second and the third powers of the 4-momentum $k_{\mu }$ in their
Fourier transforms. Aiming at the local limit, we may admit that the
effective Lagrangian $\mathfrak{L}$ depends only on (the relativistic and
gauge invariant combinations of) the fields $F_{\rho \sigma }.$ Moreover, as
long as constant fields are concerned all such combinations may be expressed
as functions of the two field invariants $\mathfrak{F}=\frac{1}{4}F_{\rho
\sigma }F^{\rho \sigma }$ and $\mathfrak{G}=\frac{1}{4}F^{\rho \sigma }%
\tilde{F}_{\rho \sigma }\mathbf{,}$ where the dual field tensor is defined
as $\tilde{F}_{\rho \sigma }=\frac{1}{2}\epsilon _{\rho \sigma \lambda
\kappa }F^{\lambda \kappa }$, with the completely antisymmetric unit tensor
defined in such a way that $\epsilon _{1230}=1.$ Then the variational
derivatives in (\ref{Pi}) and (\ref{Pi3}) can be calculated in terms of
derivatives of $\mathfrak{L}\left( \mathfrak{F,G}\right) $ with respect to
the field invariants reduced to the space- and time-independent external
field. Henceforth we shall be interested in the special case, where the
external field is a constant purely magnetic field in a certain class of
reference frames, called special below. Since in other Lorentz frames the
electric field is also present we refer to this case as magnetic-like. The
invariant conditions that specialize the magnetic-like case are $\mathfrak{F}%
>0,$ $\mathfrak{G}=0.$ Once the invariant $\mathfrak{G}$ is a pseudoscalar,
the Lagrangian of a P-invariant theory, QED included, may contain it only in
an even power. Hence all the odd derivatives of $\mathfrak{L}\left( 
\mathfrak{F,G}\right) $ with respect to it disappear after being reduced to
the external magnetic-like field: 
\begin{align}
\left. \frac{\partial \mathfrak{L}(\mathfrak{F},\mathfrak{G})}{\partial 
\mathfrak{G}}\right\vert _{F=\mathcal{F},\mathfrak{G=}0}& =\left. \frac{%
\partial ^{2}\mathfrak{L}(\mathfrak{F},\mathfrak{G})}{\partial \mathfrak{G}%
\partial \mathfrak{F}}\right\vert _{F=\mathcal{F},\mathfrak{G=}0}=0  \notag
\\
\left. \frac{\partial ^{3}\mathfrak{L}(\mathfrak{F},\mathfrak{G})}{\partial 
\mathfrak{G}\partial \mathfrak{F}^{2}}\right\vert _{F=\mathcal{F},\mathfrak{%
G=}0}& =\left. \frac{\partial ^{3}\mathfrak{L}(\mathfrak{F},\mathfrak{G})}{%
\partial \mathfrak{G}^{3}}\right\vert _{F=\mathcal{F},\mathfrak{G=}0}=0.
\label{oddder}
\end{align}%
We calculate (\ref{Pi}) and (\ref{Pi3}) in Appendix using the rule 
\begin{equation}
\frac{\delta F_{\alpha \beta }(z)}{\delta A^{\mu }(x)}=\left( \eta _{\mu
\beta }\frac{\partial }{\partial z^{\alpha }}-\eta _{\mu \alpha }\frac{%
\partial }{\partial z^{\beta }}\right) \delta ^{4}(x-z),  \label{rule}
\end{equation}%
(understood as integrated over $z$\ with any function of $z$)\ by repeatedly
applying the relation%
\begin{equation}
\frac{\delta \Gamma }{\delta A^{\mu }(x)}=\int \left[ \frac{\partial 
\mathfrak{L}(\mathfrak{F}(z),\mathfrak{G}(z))}{\partial \mathfrak{F}(z)}%
F_{\alpha \mu }(z)+\frac{\partial \mathfrak{L}(\mathfrak{F}(z),\mathfrak{G}%
(z))}{\partial \mathfrak{G}(z)}\tilde{F}_{\alpha \mu }(z)\right] \frac{%
\partial }{\partial z_{\alpha }}\delta ^{4}(x-z)\mathrm{d}^{4}z
\label{firstder}
\end{equation}%
and reducing the final results onto the external field. Then, taking eqs. (%
\ref{oddder}) into account and using the notations

\begin{align*}
\mathfrak{L_{F}} & =\left. \frac{\mathrm{d}\mathfrak{L}(\mathfrak{F},0))}{%
\mathrm{d}\mathfrak{F}}\right\vert _{F=\mathcal{F}},\ \ \mathfrak{L_{FF}}%
=\left. \frac{\mathrm{d}^{2}\mathfrak{L}(\mathfrak{F},0)}{\mathrm{d}%
\mathfrak{F}^{2}}\right\vert _{F=\mathcal{F}},\text{ }\mathfrak{L_{GG}}%
=\left. \frac{\partial^{2}\mathfrak{L}(\mathfrak{F},\mathfrak{G})}{\partial%
\mathfrak{G}^{2}}\right\vert _{F=\mathcal{F},\mathfrak{G}=0}, \\
\mathfrak{L_{FFF}} & \mathfrak{=}\left. \frac{\mathrm{d}^{3}\mathfrak{L}(%
\mathfrak{F},0)}{\mathrm{d}\mathfrak{F}^{3}}\right\vert _{F=\mathcal{F}}%
\mathfrak{,}\text{ \ \ \ }\mathfrak{L_{FGG}}=\frac{\mathrm{d}}{\mathrm{d}%
\mathfrak{F}}\left. \frac{\partial^{2}\mathfrak{L}(\mathfrak{F},\mathfrak{G})%
}{\partial\mathfrak{G}^{2}}\right\vert _{F=\mathcal{F},\mathfrak{G}=0},
\end{align*}
all with $F_{\mu\nu}=\mathcal{F}_{\mu\nu}$ substituted (hence, from now on, $%
\mathfrak{F}=\frac{1}{4}\mathcal{F}_{\rho\sigma}\mathcal{F}^{\rho\sigma}>0$
and $\mathfrak{G}=\frac{1}{4}\mathcal{F}^{\rho\sigma}\mathcal{\tilde{F}}%
_{\rho\sigma}=0)\mathbf{,}$ we get for the second-

\begin{align}
& \Pi _{\mu \tau }^{\mathrm{IR}}(x-y)=\mathfrak{L_{F}}\left( \frac{\partial
^{2}}{\partial x^{\tau }\partial x^{\mu }}-\eta _{\mu \tau }\square \right)
\delta ^{4}(x-y)-  \notag \\
& -\left\{ \mathfrak{L_{FF}}\right. \mathcal{F}_{\alpha \mu }\mathcal{F}%
_{\beta \tau }+\text{ }\mathfrak{L_{GG}}\mathcal{\tilde{F}}_{\alpha \mu
}\left. \mathcal{\tilde{F}}_{\beta \tau }\right\} \frac{\partial }{\partial
x_{\alpha }}\frac{\partial }{\partial x_{\beta }}\delta ^{4}(x-y).\quad
\label{secrank2}
\end{align}%
and for the third-rank tensor in the infrared limit%
\begin{align}
& \Pi _{\mu \tau \sigma }^{\mathrm{IR}}(x-y,x-u)=  \notag \\
& =-\mathcal{O}_{\mu \tau \sigma \alpha \beta \gamma }\frac{\partial }{%
\partial x_{\alpha }}\left( \left( \frac{\partial }{\partial x_{\beta }}%
\delta ^{4}(y-x)\right) \left( \frac{\partial }{\partial x_{\gamma }}\delta
^{4}(x-u)\right) \right) ,  \label{thirdrank2}
\end{align}%
where

\begin{align}
& \mathcal{O}_{\mu \tau \sigma \alpha \beta \gamma }=\mathfrak{L_{GG}}\left[ 
\widetilde{\mathcal{F}}_{\gamma \sigma }\epsilon _{\alpha \mu \beta \tau }+%
\widetilde{\mathcal{F}}_{\alpha \mu }\epsilon _{\beta \tau \gamma \sigma }+%
\widetilde{\mathcal{F}}_{\beta \tau }\epsilon _{\alpha \mu \gamma \sigma }%
\right] +  \notag \\
& +\mathfrak{L_{FF}}\left[ \left( \eta _{\mu \tau }\eta _{\alpha \beta
}-\eta _{\mu \beta }\eta _{\alpha \tau }\right) \mathcal{F}_{\gamma \sigma }+%
\mathcal{F}_{\alpha \mu }\left( \eta _{\tau \sigma }\eta _{\gamma \beta
}-\eta _{\beta \sigma }\eta _{\gamma \tau }\right) +\mathcal{F}_{\beta \tau
}\left( \eta _{\mu \sigma }\eta _{\gamma \alpha }-\eta _{\alpha \sigma }\eta
_{\gamma \mu }\right) \right]   \notag \\
& +\mathfrak{L_{FGG}}\left[ {\mathcal{F}}_{\alpha \mu \text{ }}\widetilde{%
\mathcal{F}}_{\beta \tau }\widetilde{\mathcal{F}}_{\gamma \sigma }+%
\widetilde{\mathcal{F}}_{\alpha \mu }\mathcal{F}_{\beta \tau }\widetilde{%
\mathcal{F}}_{\gamma \sigma }\text{\emph{+}}\widetilde{\mathcal{F}}_{\alpha
\mu }\widetilde{\mathcal{F}}_{\beta \tau }\mathcal{F}_{\gamma \sigma }\right]
\text{ }+\mathfrak{L_{FFF}}\mathcal{F}_{\alpha \mu }\mathcal{F}_{\beta \tau }%
\mathcal{F_{\gamma \sigma }}.\mathcal{\ }\text{\ }  \label{sixrank}
\end{align}%
(The reader may consult Appendix  for detailed calculations.) This tensor
turns to zero, when there is no external field, $\mathcal{F}=0,$ in
agreement with the Furry theorem. The two transversality conditions (\ref%
{transvers}) for (\ref{secrank2}) are provided by that the matrix in the
brackets is antisymmetric under each permutation $\mu \Longleftrightarrow
\alpha $ and \ $\tau \iff \beta ,$ while the first term in (\ref{secrank2})
is transverse explicitly$.$ The three transversality conditions (\ref%
{transvers}) for (\ref{thirdrank2}) are provided by that matrix (\ref%
{sixrank}) is antisymmetric under each permutation $\mu \Longleftrightarrow
\alpha ,$\ $\tau \iff \beta $ and $\sigma \iff \gamma .$ Thanks to the two
latter antisymmertries, by using (\ref{thirdrank2}), (\ref{sixrank}) in (\ref%
{nonlincur}) we obtain for the nonlinearly induced current the expression

\begin{equation}
j_{\mu }^{\text{nl}}(x)=\frac{1}{8}\mathcal{O}_{\mu \tau \sigma \alpha \beta
\gamma }\frac{\partial }{\partial x_{\alpha }}\left( f^{\beta \tau
}f^{\gamma \sigma }\right) ,  \label{nonlincur2}
\end{equation}%
that includes only the field intensity tensors $f^{\beta \tau }=\frac{%
\partial }{\partial x_{\beta }}a^{\tau }(x)-\frac{\partial }{\partial
x_{\tau }}a^{\beta }(x)$. Therefore, the nonlinearly induced current is
gauge-invariant: it depends only on field intensities and, besides, it is
conserved, $\frac{\partial }{\partial x_{\mu }}j_{\mu }^{\text{nl}}(x)=0$
due to the first antisymmetry $\mu \Longleftrightarrow \alpha .$

We have to approach the nonlinear set (\ref{a}), (\ref{nonlincur}) by
looking for its solution in a power series in the field $a^{\lambda}(x).$
Within the first iteration, to which we shall as a matter of fact confine
ourselves, we substitute the linear approximation to the solution of
equation (\ref{a})%
\begin{equation}
a_{\nu}^{\text{lin}}(x)=\int d^{4}x^{\prime}D_{\nu\rho}(x-x^{\prime})j^{\rho
}(x^{\prime})  \label{propagator2}
\end{equation}
for $a(x)$ into (\ref{nonlincur}). In other words, we should use the
electromagnetic field $\ f_{\beta\tau}=$ $f_{\beta\tau}^{\text{lin}}=\frac{%
\partial}{\partial x^{\beta}}a_{\tau}^{\text{lin}}(x)-\frac{\partial }{%
\partial x^{\tau}}a_{\beta}^{\text{lin}}(x)$ linearly produced by the source 
$j_{\mu}(x)$ in the expression for the nonlinearly induced current (\ref%
{nonlincur2}).

\section{Magnetic field of a static charge at rest in external magnetic field%
}

We are in a position to start studying the nonlinear effect of production of
a magnetic field by a static charge at rest in a constant and homogeneous
external magnetic field in a special frame. The linear effect of the
external magnetic field on the electrostatic field of a charge was studied
earlier (beyond the infrared approximation) in \cite{shabus, sad, mache}.

In this frame the external magnetic field is defined as $B_{i}=(1/2)\epsilon
_{ijk}\mathcal{F}_{jk}=\widetilde{\mathcal{F}}_{i0},$ $B=|\mathbf{B}|,$
while the external electric field disappears $E_{i}=\mathcal{F}_{0i}=0.$ The
roman indices span the \textbf{3D} subspace in this reference frame, $%
\epsilon _{ijk}$ is the fully antisymmetric tensor, $\epsilon _{123}=1.$

Consider now a static charge given in that frame by the 4-current $j_{\mu
}(x)=$ $j_{\mu }(\boldsymbol{x})\delta _{\mu 0}$. In the linear
approximation $\left( \ref{propagator2}\right) ,$ naturally, only an
electrostatic field is generated in that frame. Hence, the components with $%
\alpha =\beta =\gamma =0,$ $\tau ,$ $\sigma \neq 0$ do not contribute to (%
\ref{nonlincur2}), so we need only the components%
\begin{eqnarray}
\mathcal{O}_{i00jmn} &=&-\mathcal{F}_{ji}\left[ \text{ }\delta _{mn}%
\mathfrak{L_{FF}}-\text{ }\widetilde{\mathcal{F}}_{n0}\widetilde{\mathcal{F}}%
_{m0}\mathfrak{L_{FGG}}\right] +\left[ \widetilde{\mathcal{F}}_{m0}\epsilon
_{jin0}+\widetilde{\mathcal{F}}_{n0}\epsilon _{jim0}\right] \mathfrak{L_{GG}=%
}  \notag \\
&=&\epsilon _{ijk}B_{k}\left[ \text{ }\delta _{mn}\mathfrak{L_{FF}}-\text{ }%
B_{n}B_{m}\mathfrak{L_{FGG}}\right] +\left[ B_{m}\epsilon
_{jin}+B_{n}\epsilon _{jim}\right] \mathfrak{L_{GG}}  \label{19}
\end{eqnarray}%
in $\left( \ref{nonlincur2}\right) $ (and the ones obtained from (\ref{19})
by permutations between the second and the fifth, and between the third and
the sixth indices),\ while $\mathcal{O}_{000jmn}=0$ according to (\ref%
{sixrank})$.$ Therefore $j_{0}^{\text{nl}}(\boldsymbol{x})=0,$ i.e. there is
no nonlinear (quadratic) correction to the static charge within the current
quadratic approximation: the induced current $\left( \ref{nonlincur2}\right) 
$ is purely spacial:%
\begin{align}
& j_{i}^{\text{nl}}(\boldsymbol{x})=\frac{1}{2}\mathcal{O}_{i00jmn}\frac{%
\partial }{\partial x_{j}}\left( f_{m0}^{\text{lin}}f_{n0}^{\text{lin}%
}\right) =  \notag \\
& =\frac{1}{2}\left( \boldsymbol{\nabla \times B}\right) _{i}\left[ 
\mathfrak{L_{FF}}\boldsymbol{\mathcal{E}}^{2}-\mathfrak{L_{FGG}}\left( 
\boldsymbol{B\mathcal{E}}\right) ^{2}\right] -\mathfrak{L_{GG}}\left( 
\boldsymbol{\nabla \times \mathcal{E}}\right) _{i}\left( \boldsymbol{B%
\mathcal{E}}\right) ,  \label{nonlincur3}
\end{align}%
where $\mathcal{E}_{n}=\mathcal{E}_{n}(\boldsymbol{x})=f_{0n}^{\text{lin}}=%
\frac{-\partial }{\partial x_{n}}a_{0}^{\text{lin}}(\boldsymbol{x})$ is the
time-independent electric field, linearly produced following Eq.(\ref%
{propagator2}), and the differential operator $\boldsymbol{\nabla }$ acts on
everything to the right of it. The magnetic field strength $\boldsymbol{h}%
\mathbf{(}\boldsymbol{x}\mathbf{)}$ generated by this current according to
the Maxwell equation $\boldsymbol{\nabla }\mathbf{\times }\boldsymbol{h}%
\mathbf{(}\boldsymbol{x}\mathbf{)=}$ $\boldsymbol{j}^{\text{nl}}(\boldsymbol{%
x})$ is 
\begin{equation}
h_{i}\mathbf{(}\boldsymbol{x}\mathbf{)=}\mathfrak{h}_{i}(\boldsymbol{x}%
\mathbf{)+}\nabla _{i}\Omega ,  \label{magnstrength-1}
\end{equation}%
where 
\begin{equation}
\mathfrak{h}_{i}(\boldsymbol{x}\mathbf{)=}\frac{B_{i}}{2}\left[ \mathfrak{%
L_{FF}}(\boldsymbol{\mathcal{E}}(\boldsymbol{x}))^{2}-\mathfrak{L_{FGG}}%
\left( \boldsymbol{B\mathcal{E}}(\boldsymbol{x})\right) ^{2}\right] -%
\mathcal{E}_{i}(\boldsymbol{x})\mathfrak{L_{GG}}\left( \boldsymbol{B\mathcal{%
E}}(\boldsymbol{x})\right) ,  \label{magnstrength}
\end{equation}%
because $\boldsymbol{\nabla }\mathbf{\times }\boldsymbol{\nabla }\Omega
\equiv 0,$ and the scalar function $\Omega $ should be subjected to the
Poisson equation 
\begin{equation*}
\boldsymbol{\nabla }^{2}\Omega =-\nabla _{j}\mathfrak{h}_{j}(\boldsymbol{x}%
\mathbf{)}
\end{equation*}%
to make the magnetic field $\boldsymbol{h}\mathbf{(}\boldsymbol{x}\mathbf{)}$
obey the other Maxwell equation $\boldsymbol{\nabla h}\mathbf{(}\boldsymbol{x%
}\mathbf{)=}0$. Hence, the magnetic field is the transverse part of (\ref%
{magnstrength}):%
\begin{equation}
h_{i}\mathbf{(}\boldsymbol{x}\mathbf{)=}\left( \delta _{ij}-\frac{\nabla
_{i}\nabla _{j}}{\boldsymbol{\nabla }^{2}}\right) \mathfrak{h}_{j}(%
\boldsymbol{x}\mathbf{)=}\text{ }\mathfrak{h}_{i}(\boldsymbol{x}\mathbf{)+}%
\frac{\nabla _{i}\nabla _{j}}{4\pi }\int \frac{\mathfrak{h}_{j}(\boldsymbol{y%
}\mathbf{)}}{|\boldsymbol{x}-\boldsymbol{y}|}d^{3}y.  \label{magnstrength1}
\end{equation}%
Note that the substitution of the field of a point-like charge into (\ref%
{magnstrength1}) through (\ref{magnstrength}) would cause the divergency of
the integral in (\ref{magnstrength1}) near $\mathbf{y}=0$: the present
approach fails near the point charge, since it is not applicable to its
strongly inhomogeneous field. Dealing with the point charge would require
going beyond the infrared approximation followed to in the present work.
Nevertheless, Eq. (\ref{magnstrength1}) is sound as applied to extended
charges.

Eq. (\ref{magnstrength1}) would coincide with the magnetic induction $%
\boldsymbol{b}\mathbf{(}\boldsymbol{x}\mathbf{)=}$ $\boldsymbol{\nabla\times
a}^{\text{nl}}(\boldsymbol{x}\mathbf{)}$ if the linear vacuum magnetization
effect might be neglected, i.e. if the nonlinear correction to the field in (%
\ref{a})%
\begin{equation}
a_{\text{nl}}^{\lambda}(x)=\int d^{4}yD_{~\rho}^{\lambda}(x-y)j_{\rho }^{%
\text{nl}}(y)  \label{a2}
\end{equation}
might be taken without the contribution of the linear response function (\ref%
{thirdrank2}) $\Pi_{\mu\nu}(x-x^{\prime})$ in the photon propagator $(\ref%
{propagator})$. Taking this contribution into account results in more
complicated integrals. The situation remains simple, however, when we may
disregard the anisotropy of the linear magnetic response. The inverse
magnetic permeability tensor inherent in the second-rank polarization tensor
(\ref{secrank2}) is, in the special frame, the constant tensor \cite%
{villalbachaves}

\begin{equation*}
\mu _{ij}^{-1}=\left( 1-\mathfrak{L}_{\mathfrak{F}}\right) \delta _{ij}-%
\mathfrak{L}_{\mathfrak{F}\mathfrak{F}}B_{i}B_{j},
\end{equation*}%
whose two\footnote{%
The constant background magnetic-like field makes an uniaxial medium in any
of the special frames \cite{villalbachaves}.} eigenvalues $\mu _{\perp
}^{-1}=1-\mathfrak{L}_{\mathfrak{F}},$ and $\ \mu _{\parallel }^{-1}=1-%
\mathfrak{L}_{\mathfrak{F}}-2\mathfrak{F}\mathfrak{L}_{\mathfrak{F}\mathfrak{%
F}}$ are responsible for magnetizations linearly caused by certain conserved
constant straight-linear currents flowing along the external magnetic field$%
, $ and across it, respectively (see Appendix in \cite{convexity}). In QED,
the values $\mathfrak{L}_{\mathfrak{F}}$ and $2\mathfrak{FL}_{\mathfrak{F}%
\mathfrak{F}}$ are of the order of the fine structure constant $\alpha
=1/137,$ but depend on the field $B.$ When $B$ is very large, $B\gg m^{2}/e,$
these quantities, as found from the Euler-Heisenberg one-loop effective
Lagrangian, behave as, see, $e.g.,$ \cite{Shabus2011}

\begin{equation*}
\displaystyle\mathfrak{L}_{\mathfrak{F}}\approx\frac{\alpha}{3\pi}\ln\frac {%
eB}{m^{2}},\ \ 2\mathfrak{F}\mathfrak{L}_{\mathfrak{F}\mathfrak{F}}\approx%
\frac{\alpha}{3\pi}.
\end{equation*}
So, when $\frac{eB}{m^{2}}\gg2.7,$ the contribution of $2\mathfrak{F}%
\mathfrak{L}_{\mathfrak{F}\mathfrak{F}}$ may be neglected as compared to $%
\mathfrak{L}_{\mathfrak{F}},$ and the linear magnetization becomes
isotropic, $\mu_{\perp}^{-1}=\ \mu_{\parallel}^{-1}.$ Therefore, in this
limit, we finally have for the nonlinear magnetic induction

\begin{equation}
\boldsymbol{b}\mathbf{(}\boldsymbol{x}\mathbf{)=}\left( 1-\mathfrak{L}_{%
\mathfrak{F}}\right) ^{-1}\boldsymbol{h}(\boldsymbol{x}\mathbf{)}.
\label{induction}
\end{equation}

The electric field $\boldsymbol{\mathcal{E}}$ $=-$ $\boldsymbol{\nabla }%
a_{0}^{\text{lin}}(\boldsymbol{x}\mathbf{)}$ to be substituted in (\ref%
{nonlincur3}) and (\ref{magnstrength}) is the one that is linearly produced
via Eq. (\ref{propagator2}) by a static charge distribution within the same
infrared approximation. To determine it, note that in (\ref{propagator2})
only the propagator component $D_{00}$ participates, that, in the Fourier
representation, is $D_{00}=$ ($\boldsymbol{k}^{2}-\varkappa _{2})^{-1},$
with $\varkappa _{2}$ being one (out of three) eigenvalues of the
second-rank polarization tensor (\ref{Pi}) taken in the static limit $%
k_{0}=0 $ in the special reference frame. Once the polarization tensor is
considered in its infra-red limit (\ref{secrank2}), this quantity is \cite%
{convexity, Shabus2011} $\varkappa _{2}=\boldsymbol{k}^{2}\mathfrak{L}_{%
\mathfrak{F}}-k_{\parallel }^{2}2\mathfrak{FL_{GG}.}$ Here $\boldsymbol{k}%
^{2}=\boldsymbol{k}_{\perp }^{2}+k_{\parallel }^{2},$ and $\boldsymbol{k}%
_{\perp },k_{\parallel }$ are the momentum components of the small
electromagnetic field across and along $\boldsymbol{B}$ , respectively. Now,
the calculation of (\ref{propagator2}) for the point-like charge $j_{0}(%
\boldsymbol{x})=q\delta ^{3}(\boldsymbol{x})$ results, with the use of this
propagator, in the anisotropic Coulomb law 
\begin{equation}
a_{0}^{\text{lin}}(\boldsymbol{x})=\frac{q}{4\pi }\frac{1}{\sqrt{\epsilon
_{\perp }}\sqrt{\epsilon _{\perp }x_{_{\parallel }}^{2}+\epsilon
_{_{\parallel }}\boldsymbol{x}_{\perp }^{2}}},  \label{coulomb}
\end{equation}%
where $\epsilon _{\perp }=1-\mathfrak{L}_{\mathfrak{F}}$ and \ $\epsilon
_{_{\parallel }}=1-\mathfrak{L}_{\mathfrak{F}}+2\mathfrak{F}\mathfrak{L}_{%
\mathfrak{G}\mathfrak{G}}\ $are\ eigenvalues of the dielectric tensor \cite%
{villalbachaves} $\epsilon _{ij}=\left( 1-\mathfrak{L}_{\mathfrak{F}}\right)
\delta _{ij}+\mathfrak{L}_{\mathfrak{G}\mathfrak{G}}B_{i}B_{j},$ responsible
for polarizations, linearly caused by homogeneously charged planes parallel
and orthogonal to $\boldsymbol{B,\ }$respectively, $\boldsymbol{x}_{\perp }$
and $\boldsymbol{x}_{\parallel }$ are the coordinate components across and
along $\boldsymbol{B.}$ For large magnetic field one gets the linearly
growing asymptote from the Euler-Heisenberg Lagrangian $\displaystyle2%
\mathfrak{F}\mathfrak{L}_{\mathfrak{G}\mathfrak{G}}\approx \frac{\alpha }{%
3\pi }\frac{eB}{m^{2}}.$ This means that if $\frac{eB}{m^{2}}\mathfrak{>}%
\frac{3\pi }{\alpha }$ the dielectric component $\epsilon _{_{\parallel }}$
dominates over $\epsilon _{\perp }$, i.e. the electrization becomes highly
anisotropic, in contrast to the magnetization. In this asymptotic region Eq.
(\ref{coulomb}) becomes (if we disregard the polarization in $\epsilon
_{\perp }$ by setting $\epsilon _{\perp }=1)$ the large-distance behavior of
the potential of a point charge in a strong magnetic field calculated in the
linear approximation in\textit{\ }\cite{shabus, sad} beyond the infrared
approximation of the polarization tensor. Note that Eq. (\ref{coulomb}), as
well as its high-field limit, is only valid far from the charge. In that
domain, however, it also fits any charge, with the total value $q,$
distributed over a finite region.

\section{\protect\bigskip Some numerical estimates}

To analyze the large magnetic field limits of the induced current (\ref%
{nonlincur3}), of the resulting magnetic field (\ref{magnstrength1}) and\ of
its induction (\ref{induction}) one should also bear in mind the asymptotic
behavior $\mathfrak{L_{FGG}=-}\frac{\alpha e}{3\pi m^{2}B^{3}}.$ Then it
follows from the large external magnetic field asymptotic behavior, $\frac{eB%
}{m^{2}}\mathfrak{>>}\frac{3\pi }{\alpha }$ , of the other derivatives of
the Euler-Heisenberg Lagrangian involved in (\ref{magnstrength}) that were
listed above that in this limit 
\begin{align*}
\frac{e\mathfrak{h}_{\parallel }}{m^{2}}& \sim \frac{\alpha }{6\pi }\left[
-\left( \frac{e\mathcal{E}_{\parallel }}{m^{2}}\right) ^{2}+\left( \frac{e%
\mathcal{E}_{\perp }}{m^{2}}\right) ^{2}\frac{m^{2}}{eB}\right] ,\text{ \ \ }
\\
\text{\ \ }\frac{e\mathfrak{h}_{\perp }}{m^{2}}& \sim -\frac{\alpha }{3\pi }%
\frac{e\mathcal{E}_{\parallel }}{m^{2}}\text{\ }\frac{e\mathcal{E}_{\perp }}{%
m^{2}},\text{ \ \ \ \ }\frac{eB}{m^{2}}\mathfrak{>>}\frac{3\pi }{\alpha }.
\end{align*}%
The minus sign in the first line undicates that the induced magnetic field
diminishes the external field in the large external field regime.

We may apply the results (\ref{magnstrength}), (\ref{magnstrength1}) to
small external magnetic field $(eB/m^{2})<<1,$ as well$.$ With the
Euler-Heisenberg Lagrangian density, one has in this regime: 
\begin{equation*}
\text{\ \ \ }\mathfrak{L}_{\mathfrak{FF}}\mathfrak{=}\frac{4\alpha}{45\pi }%
\left( \frac{e}{m^{2}}\right) ^{2},\text{ \ \ }\mathfrak{L_{GG}}\mathfrak{=}%
\frac{7\alpha}{45\pi}\left( \frac{e}{m^{2}}\right) ^{2},\text{ \ \ \ }%
\mathfrak{L_{FGG}=}\frac{\alpha}{315\pi}\left( \frac{e}{m^{2}}\right) ^{4}.
\end{equation*}
The third coefficient $\mathfrak{L_{FGG}}$ does not contribute in the
leading order in $(eB/m^{2})<<1$ to the estimates

\begin{align*}
\mathfrak{h}_{\parallel }& \sim B\frac{\alpha }{45\pi }\left[ 2\left( \frac{e%
\boldsymbol{\mathcal{E}}}{m^{2}}\right) ^{2}-7\left( \frac{e\mathcal{E}%
_{\parallel }}{m^{2}}\right) ^{2}\right] ,\text{ \ } \\
\text{\ \ \ }\mathfrak{h}_{\perp }& \sim -B\frac{7\alpha }{45\pi }\text{\ }%
\left( \frac{e\mathcal{E}_{\perp }}{m^{2}}\right) \left( \frac{e\mathcal{E}%
_{\parallel }}{m^{2}}\right) ,\text{ \ }\frac{eB}{m^{2}}\mathfrak{<<}1.
\end{align*}%
In this approximation we may set $\epsilon _{\perp }=\epsilon _{_{\parallel
}}=\mu _{\perp }^{-1}=\ \mu _{\parallel }^{-1}=1.$ Therefore, $\boldsymbol{h}
$ \ $=\boldsymbol{b},$ and for the electric field of a charge outside of it,
one may use here the standard Coulomb law $\boldsymbol{\mathcal{E}}=(q/4\pi
) $ $\boldsymbol{x}/|x|^{3}$ instead of (\ref{coulomb}).

Note that $\alpha /45\pi =5\cdot 10^{-5}.$ So, for the electric field value
close to Schwinger's 1.3$\cdot 10^{16}$ V/cm, the nonlinearly produced
magnetic field makes up to $3\cdot $ $10^{-4}$ of the external magnetic
field, which must be kept below Schwinger's 4.4$\cdot 10^{13}$G in this case.

\section{Conclusion}

In this paper we have found an expression for the magnetic field $%
\boldsymbol{h}(\boldsymbol{x})$ produced by a static charge $q$ placed into
an external magnetic field $\boldsymbol{B}$, Eqs.(\ref{magnstrength}), (\ref%
{magnstrength1}). It is shown that in QED this nonlinear magneto-electric
effect, not considered before, occurs already in the simplest approximation,
where the effective Lagrangian $\mathfrak{L}$ is taken in its local limit,
and only second power of the charge $q$ and/or of its electric field $%
\boldsymbol{\mathcal{E}}(\boldsymbol{x})$ are kept. As for the background
magnetic field $\boldsymbol{B}$, to reveal the effect suffice it to take it
into account in the linear approximation, $\sim B$, although magnetic field $%
B$\ of arbitrary magnitude is included in our result, as well. The final
formulas depend on the first three derivatives of the effective action $%
\mathfrak{L}$ with respect to the external field invariants, which complies
with the fact that, minimally, diagrams with three photon legs are
responsible for the effect in the given approximation.

\bigskip

The results are model-independent and relate not only to QED, but also to
any nonlinear electrodynamics provided the standard postulates of
U(1)-gauge-, Lorentz-, translation-, C-, P-, T- invariances are respected.
When applying them to QED we take the Euler-Heisenberg Lagrangian for $%
\mathfrak{L}$ to estimate the regimes of weak and strong $\boldsymbol{B.}$
In QED all electromagnetic fields appear in ratios to the Schwinger
characteristic value $m^{2}/e$ of 4.4$\cdot 10^{13}$ cgse units. The
nonlinear magneto-electric effect we are reporting on is efficient, if the
electric field of a charge is comparable, but still smaller than $m^{2}/e$.
Such fields take place near atomic nuclei and at the surface of a strange
quark star. Strange quark stars can be strongly magnetized, besides \cite%
{AAA}. When the Schwinger value is exceeded by the electric field, the
nonlinearity can no longer be treated via the power expansion (\ref{cubic})
, and also electron-positron pair creation from the vacuum must be taken
into account.

\section*{Acknowledgements}

Authors are indebted to T. Adorno and C. Costa Lopes for correcting some
equations. D. Gitman thanks FAPESP and CNPq for permanent support and
Russian Ministry of Science for partial support under Grant No.
14.B37.21.0911. A. Shabad acknowledges the support of FAPESP, Processo
2011/51867-9, and of RFBR under the Project 11-02-00685-a. He also thanks
USP for kind hospitality extended to him during his stay in S$\tilde{\mathrm{%
a}}$o Paulo, Brazil, where this work was fulfilled.

\section*{Appendix}

The second variational derivative of the local effective action is%
\begin{align}
& \frac{\delta ^{2}\Gamma }{\delta A^{\mu }(x)\delta A^{\tau }(y)}=\int 
\mathrm{d}^{4}z\left\{ \frac{\partial \mathfrak{L}(\mathfrak{F}(z),\mathfrak{%
G}(z))}{\partial \mathfrak{F}(z)}\left( \eta _{\mu \tau }\eta _{\alpha \beta
}-\eta _{\mu \beta }\eta _{\alpha \tau }\right) \right. +  \notag \\
& +\frac{\partial \mathfrak{L}(\mathfrak{F}(z),\mathfrak{G}(z))}{\partial 
\mathfrak{G}(z)}\epsilon _{\alpha \mu \beta \tau }+\text{ \ \ \ \ \ \ \ \ } 
\notag \\
& +\frac{\partial ^{2}\mathfrak{L}(\mathfrak{F}(z),\mathfrak{G}(z))}{%
\partial (\mathfrak{F}(z))^{2}}F_{\alpha \mu }(z)F_{\beta \tau }(z)+\frac{%
\partial ^{2}\mathfrak{L}(\mathfrak{F}(z),\mathfrak{G}(z))}{\partial (%
\mathfrak{G}(z))^{2}}\tilde{F}_{\alpha \mu }(z)\tilde{F}_{\beta \tau }(z)+ 
\notag \\
& +\left. \frac{\partial ^{2}\mathfrak{L}(\mathfrak{F}(z),\mathfrak{G}(z))}{%
\partial \mathfrak{F}(z)\partial \mathfrak{G}(z)}\left[ {F}_{\alpha \mu }(z)%
\tilde{F}_{\beta \tau }(z)+\tilde{F}_{\alpha \mu }(z){F}_{\beta \tau }(z)%
\right] \right\} \left( \frac{\partial }{\partial z_{\alpha }}\delta
^{4}(x-z)\right) \left( \frac{\partial }{\partial z_{\beta }}\delta
^{4}(y-z)\right) .\quad   \label{secdir}
\end{align}%
(It was convenient to exploit the relation 
\begin{equation*}
\frac{\delta \tilde{F}_{\alpha \beta }(z)}{\delta A^{\mu }(x)}=\epsilon
_{\alpha \beta \gamma \mu }\frac{\partial }{\partial x_{\gamma }}\delta
^{4}(x-z)
\end{equation*}%
following from (\ref{rule})). 

After reduced to the constant and homogeneous magnetic-like external field
this gives the second-rank polarization tensor (\ref{secrank2}). The latter
can be also used when the external field is "crossed", i.e. when $\mathfrak{G%
}=0,$ $\mathfrak{F}=0,$ but $\mathcal{F}^{\rho \sigma }\neq 0$ (electric and
magnetic field vectors are mutually orthogonal and equal in length in every
Lorentz frame)$.$

The next derivative of (\ref{secdir}) is

\begin{align}
& \frac{\delta^{3}\Gamma}{\delta A^{\mu}(x)\delta A^{\tau}(y)\delta
A^{\sigma}(u)}=\text{\emph{\ \ \ \ \ \ \ \ \ \ \ \ \ \ \ \ \ \ \ \ \ \ }} 
\notag \\
& \int\mathrm{d}^{4}z{\Huge \{}\left(
\eta_{\mu\tau}\eta_{\alpha\beta}-\eta_{\mu\beta}\eta_{\alpha\tau}\right) %
\left[ \frac{\partial ^{2}\mathfrak{L}(\mathfrak{F}(z),\mathfrak{G}(z))}{%
\partial\mathfrak{F}^{2}(z)}F_{\gamma\sigma}(z)+\frac{\partial^{2}\mathfrak{L%
}(\mathfrak{F}(z),\mathfrak{G}(z))}{\partial\mathfrak{G}(z)\partial\mathfrak{%
F}(z)}\tilde {F}_{\gamma\sigma}(z)\right] \frac{\partial}{\partial z_{\gamma}%
}\delta ^{4}(u-z)+  \notag \\
& +\epsilon_{\alpha\mu\beta\tau}\left[ \frac{\partial^{2}\mathfrak{L}(%
\mathfrak{F}(z),\mathfrak{G}(z))}{\partial\mathfrak{F}(z)\partial \mathfrak{G%
}(z)}F_{\gamma\sigma}(z)+\frac{\partial^{2}\mathfrak{L}(\mathfrak{F}(z),%
\mathfrak{G}(z))}{\partial(\mathfrak{G}(z))^{2}}\tilde {F}_{\gamma\sigma}(z)%
\right] \frac{\partial}{\partial z_{\gamma}}\delta ^{4}(u-z)+  \notag \\
& +\left[ \frac{\partial^{3}\mathfrak{L}(\mathfrak{F}(z),\mathfrak{G}(z))}{%
\partial(\mathfrak{F}(z))^{3}}F_{\alpha\mu}(z)F_{\beta\tau}(z)F_{\gamma%
\sigma}(z)+\frac{\partial^{3}\mathfrak{L}(\mathfrak{F}(z),\mathfrak{G}(z))}{%
\partial\mathfrak{G}(z)\partial(\mathfrak{F}(z))^{2}}F_{\alpha\mu}(z)F_{%
\beta\tau}(z)\tilde{F}_{\gamma\sigma}(z)\right] \frac{\partial}{\partial
z_{\gamma}}\delta^{4}(u-z)+  \notag \\
& +\frac{\partial^{2}\mathfrak{L}(\mathfrak{F}(z),\mathfrak{G}(z))}{\partial(%
\mathfrak{F}(z))^{2}}\left[ F_{\alpha\mu}(z)\left( \eta _{\tau\sigma}\frac{%
\partial}{\partial z^{\beta}}-\eta_{\beta\sigma}\frac{\partial}{\partial
z^{\tau}}\right) +F_{\beta\tau}(z)\left( \eta _{\mu\sigma}\frac{\partial}{%
\partial z^{\alpha}}-\eta_{\alpha\sigma}\frac{\partial}{\partial z^{\mu}}%
\right) \right] \delta^{4}(u-z)+  \notag \\
& +\tilde{F}_{\alpha\mu}(z)\tilde{F}_{\beta\tau}(z)\left[ \frac{\partial ^{3}%
\mathfrak{L}(\mathfrak{F}(z),\mathfrak{G}(z))}{\partial\mathfrak{F}%
(z)\partial(\mathfrak{G}(z))^{2}}F_{\gamma\sigma}(z)+\frac{\partial ^{3}%
\mathfrak{L}(\mathfrak{F}(z),\mathfrak{G}(z))}{\partial(\mathfrak{G}(z))^{3}}%
\tilde{F}_{\gamma\sigma}(z)\right] \frac{\partial}{\partial z_{\gamma}}%
\delta^{4}(u-z)+\quad  \notag \\
& +\frac{\partial^{2}\mathfrak{L}(\mathfrak{F}(z),\mathfrak{G}(z))}{\partial(%
\mathfrak{G}(z))^{2}}\left[ \tilde{F}_{\alpha\mu}(z)\epsilon
_{\beta\tau\gamma\sigma}+\tilde{F}_{\beta\tau}(z)\epsilon_{\alpha\mu
\gamma\sigma}\right] \frac{\partial}{\partial z_{\gamma}}\delta ^{4}(u-z)+%
\text{ \ \ \ }  \notag \\
& +\frac{\partial^{2}\mathfrak{L}(\mathfrak{F}(z),\mathfrak{G}(z))}{\partial%
\mathfrak{F}(z)\partial\mathfrak{G}(z)}[{F}_{\alpha\mu}(z)\epsilon_{\beta%
\tau\gamma\sigma}\frac{\partial}{\partial z_{\gamma}}+\tilde{F}%
_{\beta\tau}(z)\left( \eta_{\mu\sigma}\frac{\partial}{\partial z^{\alpha}}%
-\eta_{\alpha\sigma}\frac{\partial}{\partial z^{\mu}}\right) +\tilde{F}%
_{\alpha\mu}(z)\left( \eta_{\tau\sigma}\frac{\partial}{\partial z^{\beta}}%
-\eta_{\beta\sigma}\frac{\partial}{\partial z^{\tau}}\right) +  \notag \\
& +F_{\beta\tau}(z)\epsilon_{\alpha\mu\gamma\sigma}\frac{\partial}{\partial
z_{\gamma}}]\delta^{4}(u-z)+\text{ \ \ \ \ \ \ \ \ \ \ \ \ \ \ \ }  \notag \\
& +\left[ {F}_{\alpha\mu}(z)\tilde{F}_{\beta\tau}(z)+\tilde{F}_{\alpha\mu
}(z){F}_{\beta\tau}(z)\right] \left[ \frac{\partial^{3}\mathfrak{L}(%
\mathfrak{F}(z),\mathfrak{G}(z))}{\partial\mathfrak{F}^{2}(z)\partial 
\mathfrak{G}(z)}F_{\gamma\sigma}(z)+\frac{\partial^{3}\mathfrak{L}(\mathfrak{%
F}(z),\mathfrak{G}(z))}{\partial\mathfrak{F}(z)\partial \mathfrak{G}^{2}(z)}%
\tilde{F}_{\gamma\sigma}(z)\right] \text{ \ }  \notag \\
& \frac{\partial}{\partial z_{\gamma}}\delta^{4}(u-z){\Huge \}}\left( \frac{%
\partial}{\partial z_{\alpha}}\delta^{4}(x-z)\right) \left( \frac{\partial}{%
\partial z_{\beta}}\delta^{4}(y-z)\right) .   \label{de3gamma}
\end{align}
Each $z$-derivative here and in the next relation is meant to apply only to
a single $\delta$-function, the closest on the right of it.

Eq. (\ref{de3gamma}) can be used for calculating higher variational
derivatives, when needed.

After we set $\mathfrak{G=}0\mathfrak{\ }$and use (\ref{oddder}) we get for
the P-even theory 
\begin{align*}
& \left. \frac{\delta^{3}\Gamma}{\delta A^{\mu}(x)\delta A^{\tau}(y)\delta
A^{\sigma}(u)}\right\vert _{\mathfrak{G=}0}= \\
& \int\mathrm{d}^{4}z{\Huge \{}\left(
\eta_{\mu\tau}\eta_{\alpha\beta}-\eta_{\mu\beta}\eta_{\alpha\tau}\right) %
\left[ \frac{\partial ^{2}\mathfrak{L}(\mathfrak{F}(z),\mathfrak{G}(z))}{%
\partial\mathfrak{F}^{2}(z)}F_{\gamma\sigma}(z)\right] \frac{\partial}{%
\partial z_{\gamma}}\delta^{4}(u-z)+ \\
& +\epsilon_{\alpha\mu\beta\tau}\left[ \frac{\partial^{2}\mathfrak{L}(%
\mathfrak{F}(z),\mathfrak{G}(z))}{\partial(\mathfrak{G}(z))^{2}}\tilde {F}%
_{\gamma\sigma}(z)\right] \frac{\partial}{\partial z_{\gamma}}\delta
^{4}(u-z)+\text{ \ \ \ } \\
& +\left[ \frac{\partial^{3}\mathfrak{L}(\mathfrak{F}(z),\mathfrak{G}(z))}{%
\partial(\mathfrak{F}(z))^{3}}F_{\alpha\mu}(z)F_{\beta\tau}(z)F_{\gamma%
\sigma}(z)\right] \frac{\partial}{\partial z_{\gamma}}\delta ^{4}(u-z)+ \\
& +\frac{\partial^{2}\mathfrak{L}(\mathfrak{F}(z),\mathfrak{G}(z))}{\partial(%
\mathfrak{F}(z))^{2}}\left[ F_{\alpha\mu}(z)\left( \eta _{\tau\sigma}\frac{%
\partial}{\partial z^{\beta}}-\eta_{\beta\sigma}\frac{\partial}{\partial
z^{\tau}}\right) +F_{\beta\tau}(z)\left( \eta _{\mu\sigma}\frac{\partial}{%
\partial z^{\alpha}}-\eta_{\alpha\sigma}\frac{\partial}{\partial z^{\mu}}%
\right) \right] \delta^{4}(u-z)+ \\
& +\tilde{F}_{\alpha\mu}(z)\tilde{F}_{\beta\tau}(z)\left[ \frac{\partial ^{3}%
\mathfrak{L}(\mathfrak{F}(z),\mathfrak{G}(z))}{\partial\mathfrak{F}%
(z)\partial(\mathfrak{G}(z))^{2}}F_{\gamma\sigma}(z)\right] \frac{\partial }{%
\partial z_{\gamma}}\delta^{4}(u-z)+\quad \\
& +\frac{\partial^{2}\mathfrak{L}(\mathfrak{F}(z),\mathfrak{G}(z))}{\partial(%
\mathfrak{G}(z))^{2}}\left[ \tilde{F}_{\alpha\mu}(z)\epsilon
_{\beta\tau\gamma\sigma}+\tilde{F}_{\beta\tau}(z)\epsilon_{\alpha\mu
\gamma\sigma}\right] \frac{\partial}{\partial z_{\gamma}}\delta^{4}(u-z)+ \\
& +\left[ {F}_{\alpha\mu}(z)\tilde{F}_{\beta\tau}(z)+\tilde{F}_{\alpha\mu
}(z){F}_{\beta\tau}(z)\right] \left[ \frac{\partial^{3}\mathfrak{L}(%
\mathfrak{F}(z),\mathfrak{G}(z))}{\partial\mathfrak{F}(z)\partial \mathfrak{G%
}^{2}(z)}\tilde{F}_{\gamma\sigma}(z)\right] \text{ \ \ \ } \\
& \frac{\partial}{\partial z_{\gamma}}\delta^{4}(u-z){\Huge \}}\left( \frac{%
\partial}{\partial z_{\alpha}}\delta^{4}(x-z)\right) \left( \frac{\partial}{%
\partial z_{\beta}}\delta^{4}(y-z)\right) .
\end{align*}

It is understood that $\mathfrak{G}$ is set equal to zero also in the
right-hand sides of this and the next equations.

We integrate by parts with the delta-function $\delta ^{4}(x-z)$ to get%
\begin{align*}
& \left. \frac{\delta ^{3}\Gamma }{\delta A^{\mu }(x)\delta A^{\tau
}(y)\delta A^{\sigma }(u)}\right\vert _{\mathfrak{G=}0}=\text{ \ \ } \\
=-& \int \mathrm{d}^{4}z\delta ^{4}(x-z)\frac{\partial }{\partial z_{\alpha }%
}{\Huge \{}\left( \eta _{\mu \tau }\eta _{\alpha \beta }-\eta _{\mu \beta
}\eta _{\alpha \tau }\right) \left[ \frac{\partial ^{2}\mathfrak{L}(%
\mathfrak{F}(z),\mathfrak{G}(z))}{\partial \mathfrak{F}^{2}(z)}F_{\gamma
\sigma }(z)\right] + \\
& +\left[ \frac{\partial ^{3}\mathfrak{L}(\mathfrak{F}(z),\mathfrak{G}(z))}{%
\partial (\mathfrak{F}(z))^{3}}F_{\alpha \mu }(z)F_{\beta \tau }(z)F_{\gamma
\sigma }(z)\right] +\tilde{F}_{\alpha \mu }(z)\tilde{F}_{\beta \tau }(z)%
\left[ \frac{\partial ^{3}\mathfrak{L}(\mathfrak{F}(z),\mathfrak{G}(z))}{%
\partial \mathfrak{F}(z)\partial (\mathfrak{G}(z))^{2}}F_{\gamma \sigma }(z)%
\right] +\quad  \\
& +\frac{\partial ^{2}\mathfrak{L}(\mathfrak{F}(z),\mathfrak{G}(z))}{%
\partial (\mathfrak{G}(z))^{2}}\left[ \tilde{F}_{\alpha \mu }(z)\epsilon
_{\beta \tau \gamma \sigma }+\tilde{F}_{\beta \tau }(z)\epsilon _{\alpha \mu
\gamma \sigma }\right] +\epsilon _{\alpha \mu \beta \tau }\left[ \frac{%
\partial ^{2}\mathfrak{L}(\mathfrak{F}(z),\mathfrak{G}(z))}{\partial (%
\mathfrak{G}(z))^{2}}\tilde{F}_{\gamma \sigma }(z)\right] +\text{\ } \\
+& \left[ {F}_{\alpha \mu }(z)\tilde{F}_{\beta \tau }(z)+\tilde{F}_{\alpha
\mu }(z){F}_{\beta \tau }(z)\right] \left[ \frac{\partial ^{3}\mathfrak{L}(%
\mathfrak{F}(z),\mathfrak{G}(z))}{\partial \mathfrak{F}(z)\partial \mathfrak{%
G}^{2}(z)}\tilde{F}_{\gamma \sigma }(z)\right] \text{ \ }{\Huge \}} \\
& \left[ \left( \frac{\partial }{\partial z_{\gamma }}\delta
^{4}(u-z)\right) \left( \frac{\partial }{\partial z_{\beta }}\delta
^{4}(y-z)\right) \right] - \\
& -\int \mathrm{d}^{4}z\delta ^{4}(x-z)\frac{\partial }{\partial z_{\alpha }}%
\frac{\partial ^{2}\mathfrak{L}(\mathfrak{F}(z),\mathfrak{G}(z))}{\partial (%
\mathfrak{F}(z))^{2}} \\
& \left[ F_{\alpha \mu }(z)\left( \eta _{\tau \sigma }\frac{\partial }{%
\partial z^{\beta }}-\eta _{\beta \sigma }\frac{\partial }{\partial z^{\tau }%
}\right) +F_{\beta \tau }(z)\left( \eta _{\mu \sigma }\frac{\partial }{%
\partial z^{\alpha }}-\eta _{\alpha \sigma }\frac{\partial }{\partial z^{\mu
}}\right) \right] \delta ^{4}(u-z)\frac{\partial }{\partial z_{\beta }}%
\delta ^{4}(y-z).
\end{align*}%
Here and in the next relation the left-most derivative $\partial /\partial
z_{\alpha }$ acts on all functions of $z$ to the right of it, whereas each
other z-derivative acts only on the first $\delta $-function placed to the
right of it.

\bigskip The third-rank polarization tensor (\ref{thirdrank2}) is obtained
from this expression by reducing onto $z$-independent fields $F(z)$=$%
\mathcal{F}$:

\begin{align*}
\Pi _{\mu \tau \sigma }^{\text{IR}}& (x-y,x-u)=\frac{-\partial }{\partial
x_{\alpha }}{\Huge \{}\left( \eta _{\mu \tau }\eta _{\alpha \beta }-\eta
_{\mu \beta }\eta _{\alpha \tau }\right) \left[ \frac{\partial ^{2}\mathfrak{%
L}(\mathfrak{F},\mathfrak{G})}{\partial \mathfrak{F}^{2}}\mathcal{F}_{\gamma
\sigma }\right] \frac{\partial }{\partial x_{\gamma }}\delta ^{4}(u-x)+ \\
& +\epsilon _{\alpha \mu \beta \tau }\left[ \frac{\partial ^{2}\mathfrak{L}(%
\mathfrak{F},\mathfrak{G})}{\partial \mathfrak{G}^{2}}\mathcal{\tilde{F}}%
_{\gamma \sigma }\right] \frac{\partial }{\partial x_{\gamma }}\delta
^{4}(u-x)+\text{ \ \ } \\
& +\left[ \frac{\partial ^{3}\mathfrak{L}(\mathfrak{F},\mathfrak{G})}{%
\partial \mathfrak{F}^{3}}\mathcal{F}_{\alpha \mu }\mathcal{F}_{\beta \tau }%
\mathcal{F}_{\gamma \sigma }\right] \frac{\partial }{\partial x_{\gamma }}%
\delta ^{4}(u-x)+ \\
& +\frac{\partial ^{2}\mathfrak{L}(\mathfrak{F},\mathfrak{G})}{\partial 
\mathfrak{F}^{2}}\left[ \mathcal{F}_{\alpha \mu }\left( \eta _{\tau \sigma }%
\frac{\partial }{\partial x^{\beta }}-\eta _{\beta \sigma }\frac{\partial }{%
\partial x^{\tau }}\right) +\mathcal{F}_{\beta \tau }\left( \eta _{\mu
\sigma }\frac{\partial }{\partial x^{\alpha }}-\eta _{\alpha \sigma }\frac{%
\partial }{\partial x^{\mu }}\right) \right] \delta ^{4}(u-x)+ \\
& +\mathcal{\tilde{F}}_{\alpha \mu }\mathcal{\tilde{F}}_{\beta \tau }\left[ 
\frac{\partial ^{3}\mathfrak{L}(\mathfrak{F},\mathfrak{G})}{\partial 
\mathfrak{F}\partial \mathfrak{G}^{2}}\mathcal{F}_{\gamma \sigma }\right] 
\frac{\partial }{\partial x_{\gamma }}\delta ^{4}(u-x)+\quad  \\
& +\frac{\partial ^{2}\mathfrak{L}(\mathfrak{F},\mathfrak{G})}{\partial 
\mathfrak{G}^{2}}\left[ \mathcal{\tilde{F}}_{\alpha \mu }\epsilon _{\beta
\tau \gamma \sigma }+\mathcal{\tilde{F}}_{\beta \tau }\epsilon _{\alpha \mu
\gamma \sigma }\right] \frac{\partial }{\partial x_{\gamma }}\delta
^{4}(u-x)+ \\
& +\left[ \mathcal{F}_{\alpha \mu }\mathcal{\tilde{F}}_{\beta \tau }+%
\mathcal{\tilde{F}}_{\alpha \mu }\mathcal{F}_{\beta \tau }\right] \left[ 
\frac{\partial ^{3}\mathfrak{L}(\mathfrak{F},\mathfrak{G})}{\partial 
\mathfrak{F}\partial \mathfrak{G}^{2}}\mathcal{\tilde{F}}_{\gamma \sigma }%
\right] \text{ \ }\frac{\partial }{\partial x_{\gamma }}\delta ^{4}(u-x)%
{\Huge \}}\frac{\partial }{\partial x_{\beta }}\delta ^{4}(y-x).
\end{align*}%
It is meant that $\mathfrak{G}=\frac{1}{4}\mathcal{F}^{\rho \sigma }\mathcal{%
\tilde{F}}_{\rho \sigma }=0,$ and $\mathfrak{F}=\frac{1}{4}\mathcal{F}_{\rho
\sigma }\mathcal{F}^{\rho \sigma }$ here. This formula is also applicable to
the crossed field $\mathfrak{G}=0,$  $\mathfrak{F}=0,$ $\mathcal{F}^{\rho
\sigma }\neq 0.$


\begin{thebibliography}{99}
\bibitem{born} {\ M. Born and L. Infeld, Proc. Roy. Soc. A \textbf{144}, 425
(1934). }

\bibitem{Stern} A. Stern, Phys. Rev. Lett\textit{.} \textbf{100}, 061601
(2008).

\bibitem{AGSV} T.C.~Adorno, D.~M.~Gitman, A.E.~Shabad, D.V.~Vassilevich,%
\textit{\ }Phys.Rev\textit{.}\ D \textbf{84}, 085031 (2011); \textbf{84},
065003 (2011); T.C.~Adorno, D.M.~Gitman, A.E.~Shabad,\textit{\ }Phys. Rev. D 
\textbf{86}, 027702 (2012).

\bibitem{plebanski} Jerzy Pleba\'{n}ski, \emph{Lectures on Nonlinear
Electrodynamics} (Nordita, Copenhagen, 1970).

\bibitem{BerLifPit} V.B. Berestetsky, E.M. Lifshits, and L.P. Pitayevsky, 
\emph{Quantum Electrodynamics} (Nauka, Moscow, 1989; Pergamon Press Oxford,
New York, 1982)

\bibitem{Co and Usov} R.P. Negreiros, F. Weber, M. Malheiro, V. Usov, Phys.
Rev. D \textbf{80,} 083006 (2009); R. Pican\c{c}o Negreiros, I.N. Mishustin,
S. Schramm, F. Weber, Phys. Rev. D \textbf{82}, 103010 (2010);  M. Malheiro,
R. Pican\c{c}o Negreiros, F. Weber, V. Usov, Journal of Physics: Conference
Series,  \textbf{312},  042018 (2011).

\bibitem{witten} D.D. Ivanenko and D.F. Kurdgelaidze, Astrofizika \textbf{1, 
}479 (1965)\textbf{;} N. Itoh, Prog. Theor. Phys., \textbf{44}, 291 (1970);
E. Witten, Phys. Rev. D \textbf{30, } 272 (1984).

\bibitem{linares} L.P. Linares, M. Malheiro, A.R. Taurines, and M. Fiolhais,
Brazilian Journ. Phys. \textbf{36, }1391 (2006).

\bibitem{FGS} E.S. Fradkin, D.M. Gitman and S.M. Shvartsman, \emph{Quantum
Electrodynamics with Unstable Vacuum }(Springer, Berlin, 1991).

\bibitem{shabus} A.E. Shabad and V.V. Usov, Phys. Rev. Lett. \textbf{98},
180403 (2007); arXiv: 0707.3475; A.E. Shabad and V.V. Usov, Phys. Rev. D 
\textbf{77}, 025001 (2008);\emph{\textquotedblleft String-Like Electrostatic
Interaction from QED with Infinite Magnetic Field.\textquotedblright } in:
\textquotedblleft Particle Physics on the Eve of LHC" (Proc. of the 13th
Lomonosov Conference on Elementary Particle Physics, Moscow, August 2007),
Ed. A.I. Studenikin, {World Scientific, Singapore}, 392 (2009),
arXiv:0801.0115 [hep-th].

\bibitem{sad} N. Sadooghi and A. Sodeiri Jalili, Phys. Rev. D \textbf{76},
065013 (2007).

\bibitem{mache} {B.}~Machet and M.I. Vysotsky, Phys. Rev. D \textbf{83},
025022 (2011); S.I. Godunov, B. Machet, and M.I. Vysotsky, Phys. Rev. D 
\textbf{85}, 044058 (2012).

\bibitem{simonov} M.A. Andreichikov, \ B.O. Kerbikov, Yu.A. Simonov,
arXiv:1210.0227 [hep-ph].

\bibitem{adler} S.L. Adler, J.N. Bahcall, C.G. Callan, and M.N. Rosenbluth,
Phys. Rev. Lett. \textbf{25}, 1061 (1970); S.L. Adler, Ann. Phys. (N.Y.) 
\textbf{67}, 599 (1971); V.O. Papanyan and V.I. Ritus, Sov. Phys. JETP 
\textbf{34}, 1195 (1972); \textbf{38}, 879 (1974); R.J. Stoneham, J.Phys. A:
Math. Gen.\textit{, }\textbf{12, }2187 (1979).

\bibitem{harding} A.K. Harding and D. Lai, Rep. Prog. Phys. \textbf{69, }%
2631 (2006).

\bibitem{shabus2010} A.E. Shabad and V.V. Usov, Phys. Rev. D \textbf{81},
125008 ({2010)}.

\bibitem{weinberg} {S. Weinberg, \emph{The Quantum Theory of Fields},
(University Press, Cambridge, 2001}).

\bibitem{ritus} V.I. Ritus, in: \emph{Problems of Quantum Electrodynamics of
Intense Field, }Proc. P.N. Lebedev Phys. Inst. \textbf{168}, 5 (Nauka,
Moscow, 1986).

\bibitem{batalin} I.A. Batalin and A.E. Shabad, Zh. Eksp. Teor. Fiz. \textbf{%
60}, 894 (1971) [ Sov. Phys. JETP \textbf{33}, 483 (1971) ].

\bibitem{ferinsshab} E. Ferrer, V. de la Incera and A.E. Shabad, Fortsch.
Phys., \textbf{32} 6, 261 (1984).

\bibitem{villalbachaves} S. Villalba-Ch\'{a}vez and A.E. Shabad,
arXiv:1206.4491, Phys. Rev. D (accepted).

\bibitem{convexity} A.E. Shabad and V.V. Usov, arXiv: 0991.0640 [hep-th].

\bibitem{Shabus2011} A.E. Shabad and V.V. Usov, Phys. Rev. D \textbf{83},
105006 (2011).

\bibitem{AAA} B.J. Ahmedov, B.B. Ahmedov, A.A. Abdujabbarov, Ap.\&SS\textbf{%
\ 338}, 157 (2012); arXiv:1110.6586 [astro-ph.SR].
\end{thebibliography}
\end{document}